\documentclass[12pt]{article}
\usepackage{amsfonts}
\usepackage{amssymb}   
\usepackage{cite}
\def\hybrid{\topmargin -20pt    \oddsidemargin 0pt
        \headheight 0pt \headsep 0pt
        \textwidth 6.25in       
        \textheight 9.5in       
        \marginparwidth .875in
        \parskip 5pt plus 1pt   \jot = 1.5ex}

\hybrid

\newcommand{\cM}{{\cal M}}

\renewcommand{\L}{\mathcal{L}}
\newcommand{\M}{\mathcal{M}}
\newcommand{\ax}{\alpha}
\newcommand{\bx}{\beta}
\newcommand{\cx}{\gamma}
\newcommand{\dx}{\delta}

\newcommand{\Sx}{\Sigma}
\newcommand{\Lx}{\Lambda}
\newcommand{\Ox}{\Omega}
\newcommand{\Dx}{\Delta}
\newcommand{\Gx}{\Gamma}

\newcommand{\V}{{\cal V}}
\newcommand{\VE}{{\cal V}_{\rm E}}
\newcommand{\C}{{\hat B}}

\newcommand{\be}{\begin{equation}}
\newcommand{\ee}{\end{equation}}
\newcommand{\bea}{\begin{eqnarray}}
\newcommand{\eea}{\end{eqnarray}}
\newcommand{\ba}{\begin{array}}
\newcommand{\ea}{\end{array}}
\newcommand{\bt}{\begin{tabular}}
\newcommand{\et}{\end{tabular}}
\newcommand{\bc}{\begin{center}}
\newcommand{\ec}{\end{center}}

\newcommand{\f}{\frac{1}{2}}
 
\def\theequation{\arabic{section}.\arabic{equation}}
\def\nn{\nonumber}

\begin{document}
\begin{titlepage}
\begin{center}

\hfill hep-th/0110187\\

\vskip 3cm
{\large \bf  Heterotic String Theory with Background Fluxes}\footnote{Work
supported by:
DFG -- The German Science Foundation,
GIF -- the German--Israeli
Foundation for Scientific Research,
European RTN Program HPRN-CT-2000-00148 and the
DAAD -- the German Academic Exchange Service.}

\vskip .5in

{\bf Jan Louis
and Andrei Micu\footnote{email: 
{\tt j.louis@physik.uni-halle.de, micu@physik.uni-halle.de}} }  \\

\vskip 0.8cm
{\em Fachbereich Physik, Martin-Luther-Universit\"at Halle-Wittenberg,\\
Friedemann-Bach-Platz 6, D-06099 Halle, Germany}

\end{center}

\vskip 2.5cm

\begin{center} {\bf ABSTRACT } \end{center}
We discuss compactifications of the heterotic string 
in the presence of background fluxes.
Specifically we consider compactifications on $T^6, T^5,
K3\times T^2$ and $K3\times S^1$ for which we derive 
the bosonic sector of the low energy effective action.
The consistency with the corresponding gauged supergravities
is demonstrated.

\vfill

October 2001
\end{titlepage}


\setcounter{equation}{0}\section{Introduction}

Ever since its invention the heterotic string \cite{GHMR}
has been studied as a possible candidate for
a unified particle theory. 
In particular, its ground states with four space-time dimensions
($d=4$) and $N=1$ supersymmetry 
have been of special interest due to their
phenomenological attraction \cite{CHSW}.
Such vacua arise, for example, as compactifications 
of the ten-dimensional heterotic string on Calabi-Yau
threefolds. However, in spite of the attractive phenomenological 
properties
there remain a few important
unsolved problems such as a viable mechanism for
hierarchical supersymmetry breaking and the
lifting of the vacuum degeneracy.

Deformations of such vacua 
where non-trivial background fluxes of a $p$-form field strength are 
turned on 
have first been studied in the mid-eighties \cite{RW,AS1,dWS}.
These fluxes arise as the ``vacuum expectation values"
of the field strength along appropriate cycles 
of the compactification manifold.
The presence of the fluxes generically induce a potential
for the moduli scalars which
(partially) lifts the vacuum degeneracy
and breaks supersymmetry \cite{PS,JM,Bac,AGNT,TV,PM1,CKLT,GKP}. 
Such vacua are  special cases of  so called generalized dimensional
reduction introduced in \cite{SS} and further studied, for example, in
refs.\ \cite{BRGPT,LLP}. 

Apart from the phenomenological issues these generalized
compactifications have also been studied 
in connection with the various dualities which hold
among seemingly different string theories \cite{HTW,dWL}.
More specifically, ref.\ \cite{BdRE,KM}
considered generalized toroidal compactifications of the heterotic string on
$T^n$  with 
background fluxes turned on 
and showed that the perturbative 
T-duality $SO(n,n+16,{\bf Z})$ is spontaneously broken by the background
fluxes. 
The fluxes can be arranged in appropriate representations
of the T-duality group and hence 
the low energy effective action is invariant
if the fluxes are transformed accordingly.
On the other hand the non-perturbative
S-duality between the heterotic string compactified
on a four-torus $T^4$ and the type IIA string compactified
on a $K3$ manifold does not hold naively 
in the presence of background fluxes \cite{HLS}.
Similarly, the duality of the heterotic string compactified
on $K3\times T^2$ and the type IIA string compactified
on a Calabi-Yau threefold in the presence of fluxes
has been discussed in refs.\ \cite{AGNT,KK,CKKL} and in  \cite{CKKL}
it was observed 
that only for a very specific subset of
fluxes the duality seems to be intact.

The background fluxes cannot be turned on continuously in string theory since
they are quantized in units of the string scale \cite{RW}.
However, in the low energy
supergravity they do appear as continuous parameters and 
thus can be discussed as small deformations
of the well known low energy effective theories
derived for vanishing flux background.
If no charged matter multiplets are among the massless modes 
the low energy supergravity turns into a 
gauged or massive supergravity where
the fluxes play the role of 
the gauge charges or the mass parameters, respectively.
This fact -- which might not be entirely obvious a priori --
thus relates various gauged supergravities 
to appropriate string compactifications with background fluxes.
However, a complete classification is not available at present.
The induced potential generically has a dilaton instability and
no Minkowski or anti-de Sitter ground state exist.
On the other hand non-trivial (BPS-) domain wall solutions can
often be found. (See, for example, refs.\ \cite{BRGPT,LLP,HLS}).

In this paper we focus on the derivation of
the low energy effective action of generalized compactifications
and in particular their
relation with the known gauged supergravities.
More specifically in section \ref{T^n} we discuss the heterotic string
compactified on $T^n$.
In 2.1 we recall toroidal compactifications without background fluxes
while in \ref{fluxtn} we summarize the results of refs.\ \cite{KM}
when 
background fluxes are turned on.
The remaining sections 2.3 -- 2.5 then check
the consistency with the known $N=4$ gauged supergravities in $d=4,5,6$.
The analysis  for $T^6$ in section~2.3 is brief since most
of the work has already been done in ref.\ \cite{BdRE}.
We generalize their result in that we consider a larger class 
of non-trivial fluxes. 
In section \ref{T^5}
we repeat the analysis for $T^5$ compactifications and show agreement with
the gauged supergravity recently constructed in ref.\ \cite{DHZ}.
In this section (and the associated appendix B)
we give a lot more details since the result is new.
As a byproduct we also derive a different form of the bosonic action
and in particular a different form scalar potential
for the $N=4$ gauged supergravity in $d=5$.
In section~2.3 we briefly comment on the case $d=6, N=4$
where unfortunately the relevant gauged supergravity
has not been constructed.\footnote{%
The $F(4)$ gauged supergravity
coupled to matter
has recently been constructed in ref.\ \cite{ADV}.
However, this does not correspond to the gauged supergravity
which arises from a $T^4$ compactification of the heterotic string
as we will argue in section 2.3.}
However, our analysis suggests
the form of the bosonic action of this `yet to be constructed'
gauged supergravity.

In section~3 we turn to compactifications on
$K3\times T^2$. 
After a brief review of the ungauged heterotic theory
in section 3.1 we discuss fluxes along $T^2$ in section
\ref{ft2}. We derive the effective action and show that 
isometries of the vector multiplet sector are gauged.
Furthermore, the consistency of the compactification 
with $N=2$ gauged supergravity is verified.
In section \ref{fK3} we repeat the analysis for
fluxes along $K3$ and show that now isometries in
the hypermultiplet sector are gauged.
Again the effective action is shown to be consistent 
with $N=2$ gauged supergravity.
Section 3.4 considers fluxes along both $T^2$ and $K3$ which corresponds
to a simultaneous gauging in the vector- and hypermultiplet sector.
Sections 3.2--3.4 generalize some of the aspects also
discussed in ref.\ \cite{CKKL}. 
Finally in section 3.5 we study the closely related compactification on
$K3\times S^1$ to $d=5$. We sketch the derivation of the low energy effective
action and the consistency with the gauged supergravity.
Section 4 contains our conclusions

Some of the technicalities are deferred to four appendices.
In appendix A we record our notation and conventions.
Appendix \ref{N4d5} contains some details of
$N=4$ gauged supergravity in $d=5$ and 
appendix \ref{sg} contains a brief  summary of gauged
$N=2$ supergravity in $d=4$. 
In appendix \ref{apD} we derive the Killing prepotentials for the geometry
obtained in section \ref{fK3}.

\setcounter{equation}{0}\section{Toroidal compactification of the
heterotic string with background fluxes}
\label{T^n}

\subsection{No background flux -- the ungauged theory}

Let start by briefly recalling the standard toroidal compactification of the
heterotic string without any  background fluxes \cite{narain,MS,dWL}.
The heterotic string in $d=10$ has 16 unbroken supercharges
and the massless bosonic
spectrum contains the metric, an antisymmetric tensor,
the dilaton  as well as  gauge fields  transforming in the adjoint
representation of the gauge group
$E_8\times E_8$ or $SO(32)$.
Compactifications on an $n$-dimensional torus $T^n$
leave all 16 supercharges unbroken and 
the massless spectrum assembles in appropriate 
supermultiplets in $d=10-n$ space-time dimensions. 
The gravitational multiplet
features the metric $g_{\mu\nu}, \mu,\nu=0,\ldots,d-1$, 
the antisymmetric tensor $B_{\mu\nu}$,
the dilaton $\phi$ and 
$n$ graviphotons as bosonic components.
All other fields reside in vector multiplets
each containing a vector $A_\mu$ and $n$ scalar fields.

At a generic point in the field space of the scalars the non-Abelian
heterotic gauge symmetry is broken to its maximal Abelian 
subgroup $[U(1)]^{16}$ and for most parts of the paper
we are confining our attention
to this Coulomb branch of the theory. 
In addition, there are
$2n$ Abelian gauge bosons which arise
as the Kaluza-Klein gauge bosons of the metric $g_{\mu \ax}, \ax=1,\ldots,n$
and the antisymmetric tensor $B_{\mu \ax}$. Thus on the Coulomb branch
the total gauge group is $[U(1)]^{2n+16}$ of which $n$ vectors reside
in the gravitational multiplet while all other sit in
$n+16$ vector multiplets.

Apart from the dilaton there are $n(n+16)$ moduli
arising from the metric components $G_{\ax \bx}$,
the antisymmetric tensor $B_{\ax \bx}$ and 
the vectors $A^{a}_\ax, a=1,\ldots, 16$. Together they 
span the Narain moduli space \cite{narain}
\begin{equation} \label{Narain}
{\cal M}\ =\ {SO(n,n+16)\over SO(n)\times SO(n+16)}\Big/\Gamma_T\quad ,
\end{equation}
where $\Gamma_T = SO(n,n+16,{\bf Z})$ is the T-duality group.

The low energy 
effective action is the appropriate supergravity in $d=10-n$
coupled to $16+n$ Abelian vector multiplets and is given by \cite{MS}
\begin{equation} 
\label{action1}
S = \int d^d x \sqrt{-g}\, e^{-\phi} \Big\{
R + (\partial_\mu \phi)^2 + 
\frac{1}{8}  \partial_\mu M^{IJ} \partial^\mu M_{JI}
- \frac{1}{4}  M_{IJ} F^I_{\mu \nu} F^{J \mu \nu} 
- \frac{1}{12} H^2_{\mu \nu \lambda} \Big\}\ , 
\end{equation}
where
\begin{eqnarray}
\label{FHdef}
F_{\mu\nu}^I &=& \partial_{\mu}A_{\nu}^I -\partial_{\nu}A_{\mu}^I\ ,\qquad
I = 1, \ldots, 16+2n\ ,\nn\\
H_{\mu \nu \lambda} &=& \partial_{\mu} B_{\nu\lambda} 
- \frac12 \omega^{\rm YM}_{\mu \nu \lambda} - \frac12\omega^{\rm G}_{\mu \nu \lambda}
\ + \ \textrm{cyclic}
\ .
\end{eqnarray}
$\omega^{\rm YM}_{\mu \nu \lambda} = \eta_{IJ} A^I_{[\mu} F^J_{\nu \lambda]}$ 
is the 
(Abelian) Yang-Mills Chern-Simons term  while $\omega^{\rm G}$
is the gravitational Chern-Simons term which plays no role
in the following.
Throughout the paper
the indices $I,J$ run over the total number of vector fields in the 
spectrum (the graviphotons plus the gauge fields in vector multiplets)
and $\eta$ is the $(16+2n)$-dimensional square matrix
\begin{equation}
\label{eta}
\eta = \left(\begin{tabular}{ccc}
0 & ${\bf 1}_n$ & 0 \\
${\bf 1}_n$ & 0 & 0 \\
0 & 0 & ${\bf 1}_{16}$ \\
\end{tabular} \right) \ ,
\end{equation}
with signature $(n,n+16)$.
$M^{IJ}$ is the $SO(n,n+16)$ symmetric matrix of toroidal moduli
\begin{equation}
  \label{e1}
  M = \left(
    \begin{array}{ccc}
      G^{-1} & - G^{-1} \C & - G^{-1} A \\ 
      -\C^{T} G^{-1} & G+ A^T A + \C^T G^{-1} \C & A + \C^T G^{-1} A \\ 
      -A^T G^{-1} & A^T + A^T G^{-1} \C & {\bf 1}_{16} + A^T G^{-1} A\\ 
    \end{array}   
  \right)\ ,
\end{equation}
where $\C_{\ax \bx} = B_{\ax \bx} +\frac12 A_\ax^a A_\bx^a$
and  we have $M_{IJ} = \eta_{IK} M^{KL}\eta_{LJ} = M^{-1}_{IJ}$.
The action (\ref{action1}) is invariant under global $SO(n,n+16)$
transformations where the vector fields transform in the vector 
representation $A_\mu \to U A_\mu$
while the moduli transform in the symmetric tensor
representation $M\to UMU^T$.   

\subsection{Turning on background flux -- the gauged theory}
\label{fluxtn}

The next step is to turn on 
background fluxes following ref.\ \cite{KM}. Let us first
recall  the notion of ``background flux". 
All string theories have $(p-1)$-form gauge fields $A_{p-1}$ 
with a $p$-form field strength $F_p=d A_{p-1}$
in their massless spectrum.
In a compactification the flux $m_{\Gamma_p}$ of the gauge field
through a p-cycle $\Gamma_p$ is defined by the 
integral $m_{\Gamma_p} = \int_{\Gamma_p} F_p$ which 
can be non-vanishing.\footnote{%
The flux is often called `internal' when $\Gamma$ is a cycle on
the compact manifold but not in space-time. In this paper we only consider
such internal fluxes.}
If $A$ is globally defined $m_\Gamma=0$ holds by
Stokes theorem. In order to have a non-vanishing flux 
one has to add 
a term to $A$ which is only locally defined. 
This in turn modifies the field strength $F_p$
by a term which is proportional to a harmonic $p$-form $\Omega_p$ 
on the 
compactification manifold
\be
F_p = dA_{p-1} + m\,  \Omega_p\ . 
\end{equation}
$m$ is constant (in space-time and on the compact manifold)
and the background flux is obtained by
evaluating the integral 
$m_{\Gamma_p} =\int_{\Gamma_p} F= m\int_{\Gamma_p}\Omega_p$.
Generically the gauge invariance of $A_{p-1}$ insures
the consistency of this procedure. However,
whenever Chern-Simons type terms are present 
the consistency often puts a constraint on the allowed fluxes.

Let us now summarize the results of refs.\ \cite{KM}
where toroidal compactifications
with background fluxes have been discussed. 
On the Coulomb branch of
the ten-dimensional heterotic string there are $16$ two forms $F^a$
and one three-form $H$. On $T^n$ there are 
the standard one-cycles 
while two- and three-cycles are just products of
one-cycles. Consequently the  non-trivial  background  fluxes are
given by 
\begin{equation}
2 m^a_{\ax \bx} = F^a_{\ax \bx}\ , \qquad 3 \beta_{\ax \bx \cx} = H_{\ax \bx
    \cx}  \ , \quad 
\end{equation}
where $m^a_{\ax \bx}, \beta_{\ax \bx \cx}$ are constants and we follow
the convention and notation of ref.\ \cite{KM}.
In toroidal compactification additional fluxes can be turned on
due to the fact that starting in $d=8$ axionic scalars appear 
in the spectrum as metric deformations. 
These are scalar fields $a$ which only appear via
their field strength $da$ in the low energy effective action.
Consequently the action is invariant under a Peccei-Quinn (PQ)
symmetry $a \to a + {\rm const.}$ and such scalar fields can be viewed as
zero-forms. Hence it is possible to include
a non-vanishing background flux $\gamma^\ax_{\bx \cx}$
for their field strength (see \cite{KM} for more details). 
However, not all flux parameters introduced above are independent since the 
Bianchi identity $dH = - \frac12 F^a\wedge F^a$  imposes
the constraints\footnote{Strictly speaking the Bianchi identity
also includes a term $R\wedge R$ which vanishes on the torus 
but could contribute on a twisted torus. Following \cite{KM} we neglect
this term in our analysis.}
\begin{equation}
\label{T^ncon}
  m^a_{[\ax \bx} m^a_{\cx \dx]}=3\, \beta_{\epsilon [\ax
  \bx}\gamma^\epsilon_{\cx \dx]}\ . 
\end{equation}

Turning on `small' fluxes does not modify the light spectrum
but does change the action. One finds \cite{KM}
\begin{equation}
\label{eq3}
S = \int d^d x \sqrt{-g}\, e^{-\phi} \Big\{
R + (\partial_\mu \phi)^2 + 
\frac{1}{8}  \mathcal{D}_\mu M^{IJ} \mathcal{D}^\mu M_{JI}
- \frac{1}{4}M_{IJ}  F^I_{\mu \nu} 
F^{J \mu \nu} - \frac{1}{12} H^2_{\mu \nu \lambda} - \V \Big\}\, , 
\end{equation}
where
\begin{eqnarray}
\label{potKM}
\mathcal{D}_\mu M^{IJ} & = & \partial_\mu M^{IJ} - f_{KL}^I A_\mu^K M^{LJ}
- f_{KL}^J A_\mu^K M^{IL}\ ,
\\
\V & = & \frac{1}{12} M^{IL} M^{JM} M^{KN} f_{IJK}f_{LMN} - \frac{1}{4} 
M^{IL} \eta^{JM} \eta^{KN} f_{IJK}f_{LMN}\ . \nn
\end{eqnarray}
The $f_{KL}^I$ are the structure constants of a non-Abelian gauge symmetry
which is a subgroup of the isometry group
$SO(n,n+16)$ of the scalar manifold 
${\cal M}$ given in (\ref{Narain}). They are given in terms of
the flux parameters by\footnote{%
The indices on $f$ are raised and lowered with $\eta$.
Strictly speaking, they should be raised and lowered with the
Cartan metric which is not possible here since the gauge group is 
generically not semi-simple. However, due to the fact that
this group is embedded in the isometry group $SO(n,n+16)$, 
we can raise and lower the indices with $\eta$ which is 
the metric on the coset ${\cal M}$. 
(See also the discussion in \cite{KM}.)}
\be
\label{fflux}
f^a_{\alpha\beta} = 2 m^a_{\ax\bx}\ , \qquad 
f_{\ax\bx\cx} = -3 \beta_{\ax\bx\cx} \ , 
\qquad  f_{\bx\cx}^\ax = \gamma^\ax_{\bx\cx}\ .
\ee

The moduli scalars are charged under the gauge group as signaled
by the covariant derivatives in (\ref{potKM}) and a potential $\V$
is induced.
Accordingly also the field strength $F^I_{\mu \nu}$ turns into 
a non-Abelian field strength while 
$H_{\mu \nu \lambda}$ is  modified by a non-Abelian
Chern-Simons term \cite{KM}
\begin{eqnarray}
\label{FHdefmod}
F_{\mu\nu}^I &=& \partial_{\mu}A_{\nu}^I -\partial_{\nu}A_{\mu}^I
+ f^I_{JK}\, A_\mu^J A_\nu^K \ ,\nn\\
H_{\mu \nu \lambda} &=& \partial_{\mu} B_{\mu\nu} 
- \frac12 \omega^{\rm YM}_{\mu \nu \lambda} 
- \frac12 \omega^{\rm G}_{\mu \nu \lambda}\   +\ \textrm{cyclic}\ ,\\
\omega^{\rm YM}_{\mu \nu \lambda} & = &  
\eta_{IJ} A^I_{[\mu} F^J_{\nu \lambda]}  +\frac13
f_{IJK} A^I_\mu A^J_\nu A^K_\lambda  .\nn
\end{eqnarray}

Apart from the local gauge symmetry 
the action is also manifestly invariant under the T-duality
group $\Gamma_T = SO(n,n+16,{\bf Z})$ if the fluxes
transform in appropriate representations of $\Gamma_T$.
In this sense the flux parameters break the duality symmetry spontaneously.
Finally, the potential has a dilaton instability due to the $e^{-\phi}$ factor 
in (\ref{eq3}).

\subsection{The gauged supergravity of the $T^6$ compactification}
The purpose of this section is to show that the action 
(\ref{eq3}) is a special case of the most general
action of gauged $d=4,N=4$ supergravity as constructed in ref.\ \cite{dR}.
This generalizes the analysis of ref.\ \cite{BdRE}
where  a subset of fluxes was considered. More specifically
ref.\ \cite{BdRE} started from the standard heterotic string compactified
on $T^5$ without any background fluxes and performed a further $S^1$
reduction with background fluxes turned on along this $S^1$. It was shown
that the resulting bosonic action is a special case of the 
gauged supergravity of ref.\ \cite{dR}.
Here we generalize this analysis in that we show
that also for arbitrary flux parameters as considered in \cite{KM}
the action (\ref{eq3}) is a special case of 
the bosonic action constructed in  \cite{dR}.
In doing so we closely follow the procedure of ref.\ \cite{BdRE}.

In $d=4, N=4$ the vector multiplet contains 
6 scalar fields $A_\ax, \ax=1,\ldots,6$. In $T^6$ compactifications of the heterotic string
there are $16+6=22$ such
Abelian vector multiplets containing $6\cdot 22$ scalar fields
which span the coset $SO(6,22)/SO(6)\times SO(22)$.
These scalar fields are conveniently assembled in a $28\times 6$ matrix
$Z^I_\ax$ 
which satisfies the $SO(6,22)$ invariant constraint
\begin{equation}
  Z^I_\ax \eta_{IJ}Z^J_\bx = -\delta_{\ax \bx}\ .
\end{equation}
We do not recall the standard $N=4$ action in terms of the field
variables  $Z^I_\ax$ here and instead
refer the reader to the literature \cite{dR,BdRE,Wg}.\footnote{%
The reason is that  $T^6$ compactifications are
already discussed in ref.\ \cite{BdRE} and here
we only slightly generalize their analysis. Furthermore, 
in the next section when we discuss $T^5$ compactifications
we give a lot more details and they strongly resemble
the $T^6$ situation.}
In order to compare this $N=4$ action with the action
displayed in (\ref{eq3}) one has to go a different
set of coordinates defined by \cite{BdRE}
\begin{equation}\label{MZrel}
  M^{IJ} = \eta^{IJ} + 2 Z^I_\ax Z^J_\bx \delta^{\ax \bx} \ ,
\end{equation}
which are elements of $SO(6,22)$ in that they satisfy $M\eta M = \eta$.
In the string frame the $N=4$ supergravity action -- now expressed in terms of
the $M$ variables -- is indeed given by eq.\ (\ref{eq3}) 
as was shown in ref.\ \cite{BdRE}.
However, the  most general possible $N=4$ potential reads\cite{dR,BdRE}
\begin{equation}\label{eq2}
\V =  \frac{1}{12}\big(2\eta^{IL}\eta^{JM}\eta^{KN} 
       - 3\eta^{IL}\eta^{JM}M^{KN} + M^{IL}M^{JM}M^{KN}\big)
    f_{IJK}f_{LMN}\ .
\end{equation}
Exactly as in the case of toroidal compactifications the  
$f_{IJK}$ are the structure constants of the non-Abelian
gauge symmetry. The difference being that in toroidal compactifications
the $f_{IJK}$ are in turn determined by  the background fluxes
according to eq.\ (\ref{fflux}).
By inspection we see that the potential obtained in eq.\ (\ref{potKM})
corresponds to the last two terms of the potential given in (\ref{eq2}). 
What is left to show is that the first term in  (\ref{eq2})
vanishes for the specific fluxes which arise in $T^6$ compactifications
and are given in eq.\ (\ref{fflux}).
This first term in (\ref{eq2}) is  nothing but the 
contraction of the Killing form 
$K_{IJ} = f_{IK}^L f_{LJ}^K$ with $\eta$.
Due to the special form of the structure constants (\ref{fflux}),
the only non-zero components of the Killing form are
\begin{equation}
  K_{IJ}= 8 \gamma^\cx_{\ax \dx} \gamma^\dx_{\cx \bx} \ \delta^\ax_I
  \delta^\bx_J \ . 
\end{equation}
Using (\ref{eta})
it follows immediately that $\eta^{IJ} K_{JI}=0$ and hence
\begin{equation}\label{fzero}
  \eta^{IL}\eta^{JM}\eta^{KN} f_{IJK}f_{LMN} = 0\ .
\end{equation}

To summarize, we just showed that in $d=4$ the action obtained from generalized
toroidal $T^6$ compactification of the heterotic string
with non-trivial background fluxes as obtained in refs.\
\cite{BdRE,KM} is a special case of the most general
action of gauged $N=4$ supergravity constructed in 
refs.\ \cite{dR}.\footnote{It would be interesting 
to see if there are  string backgrounds where
$\eta^{IJ} K_{IJ}\neq 0$ holds such that the first term in (\ref{eq2}) 
does not vanish anymore.}

\subsection{The gauged supergravity of the $T^5$ compactification}
\label{T^5}
The next step is to redo the previous analysis for $T^5$ compactifications
and show the consistency with the gauged supergravity recently
constructed in ref.\ \cite{DHZ}.
In $d=5$ the gravitational multiplet contains
apart from the metric and the dilaton 6 vector
fields which transform in the  \textbf{5} + \textbf{1} representation
of the R-symmetry group $USp(4)\sim SO(5)$.
The singlet  plays the role of a `spectator' gauge field and 
will later on be dualized to the antisymmetric
tensor $B_{\mu\nu}$ of the heterotic string. 
In addition the spectrum features 21 vector multiplets 
each containing 5 scalar fields.
So apart from the dilaton  there are $5\cdot21$ scalars parameterizing the
coset $SO(5,21)/SO(5)\times SO(21)$.
In ref.\ \cite{DHZ} the scalar fields are written as $SO(5,21)$ matrices
$(L_I^{[\Sx \Lx]},{L_I}^{i})$ satisfying the constraints
\begin{eqnarray}
  \label{sccon}
L_{I {i}} L_J^{i}  - L_{I[\Sx \Lx]} L^{[\Sx \Lx]}_J   &=&
    \eta_{IJ}\ ,\nn\\
- {L_I}^{\Sx \Lx}L_{J\Dx \Ox} \, \eta^{IJ}& =& \delta_\Dx^{[\Sx}
 \delta_\Ox^{\Lx]} - \frac{1}{4} \Omega^{\Sx \Lx} \Omega_{\Dx \Ox}\ ,\\
 L_{I {i}}  L^{Ij} &=& \delta_i^j \ ,\nn
\end{eqnarray}
where  
$\Sx, \Lx$ are $USp(4)$ indices $\Sx, \Lx = 1,\ldots,4$
and $\Omega_{\Dx \Ox}$ is the invariant symplectic two-form
of $USp(4)$. The $L_{I[\Sx \Lx]}$ transform in the
six-dimensional representation
of $USp(4)$ while the $L_{I {i}}$ reside in the vector
representation of $SO(21)$, i.e.\ $i= 1, \ldots, 21$ 
(for more details see appendix B).

In ref.\ \cite{DHZ} it was shown that different
subgroups $K$ of $SO(5,21)$ 
can be gauged and if $K$ does not contain an Abelian factor 
-- as is the case in $T^5$ compactifications of the heterotic string --
the bosonic part of the gauged supergravity action
in the Einstein frame is given by \cite{DHZ} 
\begin{eqnarray}
  \label{act5}
  S & = & \int d^5x\sqrt{-g} \Bigg[ \frac{1}{2} R 
  - \frac{1}{2} \left(\partial_\mu \phi\right)^2
  - \frac{1}{4}
  e^{\frac{2\phi}{\sqrt 3}} M_{IJ} F^I_{\mu \nu} F^{\mu \nu J} - \frac{1}{4}
  e^{-\frac{4\phi}{\sqrt 3}}  G_{\mu \nu}G^{\mu \nu} 
  \nonumber \\
  && \quad  - \frac{1}{2} P_\mu^{{i} [\Sx \Lx]} P^\mu_{{i} [\Sx
    \Lx]}  - e^{\frac{-2 \phi}{\sqrt 3}} \V \Bigg] 
  + \frac{\sqrt 2}{8}\int d^5x\,  \eta_{IJ}  F^I_{\mu \nu}
  F^J_{\rho \sigma} a_\lambda \epsilon^{\mu \nu \rho \sigma \lambda}\ ,
\end{eqnarray}
where the `spectator' gauge field
is denoted by $a_\mu$ and its field strength by $G_{\mu\nu}$.
The kinetic terms for the scalars involve
\begin{equation}
  \label{a}
  P_\mu^{{i}\, [\Sx \Lx]}\equiv L_I^{{i}} \mathcal{D}_\mu L^{I\, [\Sx
  \Lx]}\ , 
\end{equation}
where $\mathcal{D}_\mu$ is the covariant derivative with respect to 
the gauge group and it can be found explicitly in ref.\ \cite{DHZ}.
The gauge couplings are given by\footnote{Anticipating the result
we denote the gauge couplings by $M_{IJ}$ instead of
$a_{IJ}$ used in \cite{DHZ}.} 
\begin{equation}\label{gct5}
   M_{IJ} = L_{I[\Sx \Lx]} L_J^{[\Sx \Lx]} + L_{I{i}} L_J^{i}\ ,
\end{equation}
while  the scalar potential reads
\be\label{Vfive}
\V = 2 \Big(L^{Ji} L^M_{i} {{L^K}_\Sx}^\Ox L_{I\Ox\Lx}
   L^{N\Sx\Dx} {L_{L\Dx}}^\Lx  - \frac{4}{9}\, {L^J}_{\Sx\Ox}
  {L_I}^{\Ox\Dx} {L^K}_{\Dx\Lx}
  {L^{M\Sx}}_\Xi {L_L}^{\Xi\Gx} {{L^N}_\Gx}^\Lx\Big)\, f_{JK}^I  f_{MN}^L\  . 
\end{equation}

So far we just recalled the results of ref.\ \cite{DHZ}.
Now we have to show that the action (\ref{eq3}) for $d=5$ is a special case of 
(\ref{act5}).
In particular this involves the identification
of the potential, the kinetic terms for the moduli
and the dualization of the spectator gauge field $a_\lambda$
in terms of an antisymmetric tensor $B_{\mu\nu}$.
First let us note that the matrix ${L_I}^{i}$ is the only bosonic
quantity which carries an $SO(21)$ vector index $i$. Hence it always
appears quadratically in the action and thus can
be eliminated using (\ref{sccon}).\footnote{For the fermionic part of the
  Lagrangian, the discussion is more involved and we will not get into details
  here.} We are then left with the matrices $L^I_{\Sx \Lx}$ 
describing  the scalar manifold.
As before it is convenient to define a new set of matrices 
$Z_\alpha^I$ by
\begin{equation}
  Z_\alpha^I = \frac{1}{2} {\left( \Gamma_\ax \right)_\Lx}^\Sx\, 
{{L^I}_\Sx}^\Lx \, , 
\end{equation}
where the $\Gamma^\ax$ are Euclidean 
$4\times4$ $\Gamma$-matrices. 
The conventions are summarized in appendix~B where we also show
the identities
\begin{equation}
  \label{cstz}
  Z_\ax^I Z_\bx^J \eta_{IJ} = - \delta_{\ax \bx}\ ,\qquad
 {L_I}^{\Sx \Lx} L_{J\Sx \Lx} =  Z_{\ax I} Z_{\bx J} \delta^{\ax \bx}  
\ \equiv \ Z_{IJ}\ .
\end{equation}
Inserting (\ref{cstz}) into (\ref{gct5}) and using (\ref{sccon})
we arrive at 
\begin{equation}\label{MZ}
 M_{IJ} = \eta_{IJ} + 2 Z_{IJ}   \ ,
\end{equation}
which is the five-dimensional analog of eq.\ (\ref{MZrel}).
With the help of (\ref{MZ}) we are able to rewrite the kinetic term
for the scalars in terms of the matrix $M_{IJ}$.
Starting from 
its definition (\ref{a}) and  using (\ref{sccon}), (\ref{cstz}), (\ref{cstg})
we obtain
\bea\label{PPone}
 \frac{1}{2} P_\mu ^{i \Sx \Lx} P^\mu _{i \Sx \Lx} &=& 
  \frac {1}{2} L_I^{i}
  \mathcal{D}_\mu L^{I \Sx \Lx} L_{J i } \mathcal{D}^\mu L^J_{\Sx \Lx}
 =  \frac{1}{2} \left(\eta_{IJ} + Z_{IJ}\right) \mathcal{D}_\mu Z^I_\ax
  \mathcal{D}^\mu Z^J_\bx  \delta^{\ax \bx} \nn\\
& = & - \frac{1}{2} \eta_{IJ} \eta_{KL} Z_\ax^K Z_\bx^L \mathcal{D}_\mu 
  Z_\ax^I \mathcal{D}^\mu Z_\bx^J + \frac{1}{2} \eta_{IK} \eta_{JL} Z_\ax^K Z_\ax^L
  \mathcal{D}_\mu Z_\bx^I \mathcal{D}^\mu Z_\bx^J \ .
\eea
Taking the covariant
derivative of (\ref{cstz}) one also derives
\begin{equation}\label{DZ}
  Z_\bx^J\mathcal{D}_\mu Z_\ax^I \eta_{IJ} = - Z_\ax^I\mathcal{D}_\mu Z_\bx^J
  \eta_{IJ} \ ,
\end{equation}
which when inserted into (\ref{PPone}) results in
\be\label{Pf}
 \frac{1}{2} P_\mu ^{i \Sx \Lx} P^\mu _{i \Sx \Lx}
  = - \frac{1}{4} \eta_{IK}\eta_{JL} \mathcal{D}_\mu Z^{IJ}
  \mathcal{D}^\mu Z^{KL}
= - \frac{1}{16} \mathcal{D}_\mu M^{IJ} \mathcal{D}^\mu M_{JI}
\ .
\ee
(The last equality used (\ref{MZ}) and again the fact that $\eta_{IJ}$ is
covariantly constant.)

Next we also rewrite the potential in terms of $M_{IJ}$.
Starting from (\ref{Vfive}) using (\ref{sccon}) and
(\ref{dfz}) we obtain
\begin{eqnarray}
 \V  &=&  - \frac{1}{8}  \Big[ \left( Z^{IL} + \eta^{IL} \right)
  Z_\ax^J Z_\bx^K Z_\cx^M Z_\dx^N \; Tr \!\left(\Gamma^\ax \Gamma^\bx \Gamma^\cx \Gamma^\dx
  \right) \nonumber \\ 
  & & \qquad -  \frac{1}{9} Z_\ax^I Z_\bx^J Z_\cx^K Z_\dx^L Z_\epsilon^M
  Z_\phi^N \; Tr \! 
  \left(\Gamma^\ax \Gamma^\bx \Gamma^\cx \Gamma^\dx \Gamma^\epsilon
  \Gamma^\phi \right) \Big]f_{IJK} f_{LMN}\ .
\end{eqnarray}
The traces can be computed using (\ref{GammaM})
and taking into account the antisymmetry of the
structure constants the form of the potential is found to be 
\begin{eqnarray}
  \label{V2}
  \V & = &  Z^{JM} Z^{KN} (\eta^{IL} + \frac{2}{3} Z^{IL}) f_{IJK} f_{LMN}
\nonumber \\
  & = & \frac{1}{12} \Big(2 \eta^{IL} \eta^{JM}\eta^{KN} -3 \eta^{IL}
    \eta^{JM} M^{KN} + M^{IL} M^{JM} M^{KN} \Big) f_{IJK} f_{LMN}  \ ,  
\end{eqnarray}
where the last equality used again (\ref{MZ}).
We see that the potential of the $d=5$ action can be rewritten 
exactly in the same form
as in the $d=4$ action. This result is a byproduct of our analysis.
By the exact same argument as in $d=4$ the first term in the potential 
vanishes
for the fluxes turned on in $T^5$ compactifications (\ref{fflux}).
Inserting (\ref{Pf}), (\ref{V2}) into (\ref{act5})
and performing  a Weyl rescaling of the metric we arrive at
\begin{eqnarray}
  \label{eq:act5.1}
  S & = & \frac{1}{2}\int d^5x \sqrt{-g} \, e^{-\phi} 
  \Bigg[R + \left(\partial_\mu  \phi\right)^2 
  + \frac{1}{8}  \mathcal{D}_\mu M^{IJ} \mathcal{D}^\mu
  M_{JI} - \V  \nonumber \\
  & & - \frac{1}{4}
  M_{IJ} F^I_{\mu \nu} F^{J\mu \nu} - \frac{1}{4} e^{2\phi}
  G_{\mu \nu}G^{\mu \nu}\Bigg]  + \frac{1}{8} \int d^5x\, \eta_{IJ} \,
  F^I_{\mu \nu} F^J_{\rho \sigma} a_\lambda  \epsilon^{\mu \nu\rho \sigma
  \lambda} \ .
\end{eqnarray}

The final step is to dualize the spectator gauge field $a_\lambda$ to
an antisymmetric tensor. The last term in (\ref{eq:act5.1}) will give upon
dualization the right form of the Chern-Simons correction to $H$ which on the
string theory side is dictated by anomaly cancellation.
The details can be found in appendix \ref{N4d5} with the final result being the
action of eq.\ (\ref{eq3}).

\subsection{Comments on $T^4$ compactification}

The same analysis as in the previous two sections can be carried out for $T^4$.
Unfortunately, the most general $N=2$ (16
supercharges) gauged supergravity in six dimensions has not
been constructed yet. In particular, the supergravity recently derived in ref.\ \cite{ADV}
does not include the gauged supergravity which is obtained by compactification
with fluxes. The reason is that the group which is gauged in \cite{ADV} is a
direct product of $SU(2) \subset SO(4) \subset SO(4,m)$ and a semi-simple group
$G \subset SO(m) \subset SO(4,m)$, while the group which becomes gauged
due to the presence of fluxes in the compactification on $T^4$ can not be put
in such a form.
However, one can naturally generalize the result of \cite{ADV} (or at
least the potential) to accommodate bigger groups 
 $G' \subset SO(4,m)$.\footnote{%
The precise way to do this is more involved and to give more details here
would mean to introduce the whole notation used in \cite{ADV}. The main idea
is to put the groups $SU(2)$ and $G$ on the same footing and use generic
structure constants which belong to a group $G' \supset SU(2) \times G$. We
are grateful to Marco Zagermann for clarifying this point.}
This procedure leads to a potential of the form (\ref{V2}) which, as it was
argued in the previous sections, is in agreement with the potential which
comes from compactification.

\setcounter{equation}{0}\section{$K3\times T^2$ compactification with fluxes}

\subsection{No background flux -- the ungauged theory}
Let us turn to compactifications of the heterotic string on 
$K3\times T^2$ and first briefly review the ungauged theory 
\cite{CAFP,dWKLL,AFGNT,LF,N2}.
This class of heterotic string vacua
has four flat Minkowski dimensions and
leaves 8 supercharges unbroken. Thus the low energy effective
theory is an $N=2$ supergravity coupled to 
 $n_V$ vector- and $n_H$ hypermultiplets.\footnote{%
Strictly speaking
the heterotic dilaton resides in a vector-tensor multiplet \cite{dWKLL}
but this subtlety will play no role here. A short review about $N=2$
supergravity is assembled in appendix~C.}
The precise number of vector- and hypermultiplets
depends on the details compactifications and in particular
on the specific solution of the constraint
\be\label{K3cons}
\int_{K3} dH = \int_{K3} (tr R\wedge R -  tr F\wedge F) 
= 24 - \int_{K3} tr F\wedge F = 0\ .
\ee
(\ref{K3cons}) implies that the gauge bundle on $K3$ is necessarily non-trivial
with instanton numbers which have to add up to 24. This in turn breaks
the heterotic gauge group and introduces a set of bundle
moduli fields which parameterize the embedding of
the instantons in the gauge bundle.
Furthermore, depending on the embedding of the instantons
charged hypermultiplets appear among the massless modes.
In the following  
we do not need to discuss the generic situation since we will only
consider fluxes for (Abelian) gauge fields under which no
hypermultiplet is charged. Hence in this section we discuss 
a spectrum of 
$n_V$ Abelian vector- and $n_H$ gauge neutral hypermultiplets.

The low energy effective action can be derived by 
performing the standard Kaluza-Klein reduction 
of the ten-dimensional heterotic theory.
The final action is known for some time \cite{CAFP,dWKLL,AFGNT,LF}  
but for later use we present some details in the following.
After an appropriate 
Weyl rescaling and the dualization of the antisymmetric
tensor $B_{\mu\nu}$ to an axionic scalar $a$, one finds 
\bea
  \label{an2}
  S &=&  2 \int d^4 x \sqrt{-g} \;  \Big[ \f R 
+ \frac{1}{4}\, I_{IJ} F^I_{\mu \nu} F^{J\mu \nu}  
+ \frac{1}{8} \, R_{IJ} F^I_{\mu \nu} F^J_{\rho\lambda}
\epsilon^{\mu \nu \rho \lambda} \nn\\
&&\quad + \frac{\partial_\mu s\partial^\mu \bar{s}}{(s - \bar{s})^2}  
  + \frac{1}{16}\,  \partial_\mu M^{IJ} \partial^\mu M_{JI} 
  - h_{uv} \partial_\mu q^u \partial^\mu q^v
  \Big] \; ,
\eea
where the $q^u, u = 1,\ldots,4n_h$
denote the scalars in the hypermultiplets  and $h_{uv}$
is their quaternionic metric. Due to the presence of the bundle moduli 
which reside in hypermultiplets this metric is largely unknown.
$N=2$ supergravity constrains the metric to be quaternionic and 
furthermore in $K3$ compactifications 
the moduli space of the hypermultiplets 
${\cal M}_H$ has a submanifold
spanned by the moduli of the $K3$ surface which is given by \cite{NS}
\begin{equation}\label{K3sub}
{\cal M}_H\supset {\cal M}_{K3} = {SO(4,20)\over SO(4)\times SO(20)}\ \ .
\end{equation}

The variable $s$ in (\ref{an2}) denotes the heterotic dilaton 
which is defined as
\be
s\ =\ \frac{a}2  - \frac{i}{2}\, e^{-\phi}\ ,
\ee
and the gauge couplings are found to be
\be\label{IM}
I_{IJ} =  \frac{(s - \bar{s})}{2 \imath}\ M_{IJ}
\ ,\qquad R_{IJ}  =  - \frac{s + \bar{s}}{2}\
\eta_{IJ}\ ,
\ee
where the matrix $M_{IJ}$ is defined in (\ref{e1}).
The dilaton $s$ together with the toroidal moduli of $M_{IJ}$
form the moduli of the vector multiplets and $N=2$ supergravity requires that
they span a special K\"ahler manifold ${\cal M}_V$ (for more details see 
appendix~C). 
For the heterotic string compactified on $K3\times T^2$ this known to be
\cite{CAFP,dWKLL,AFGNT,LF,N2}
\begin{equation}\label{SKcoset}
{\cal M}_V\ =\ {SU(1,1)\over U(1)}\otimes 
{SO(2,n_V-1)\over SO(2)\times SO(n_V-1)}\ ,
\end{equation}
where 
the  ${SU(1,1)\over U(1)}$ factor
is spanned by the heterotic dilaton multiplet $s$. 
Indeed, one can find appropriate
complex K\"ahler coordinates by  the transformation  \cite{KK}
\begin{eqnarray}
  \label{e11}
  A_1^a & = & \sqrt 2\ \frac{n^a - \bar{n}^a}{u - \bar{u}} \ ,\qquad\qquad
  A_2^a = \sqrt 2 \ \frac{\bar{u} n^a - u \bar{n}^a}{u - \bar{u}}  \ , \nn \\  
  B_{12} & = & \frac{1}{2} \left[(t + \bar{t}) - 
    \frac{(n + \bar{n})^a(n - \bar{n})^a}{u - \bar{u}} \right]\ , \nn\\
    \sqrt{G} &=&  -\frac{\imath}{2} \left[ 
(t - \bar{t}) -\frac{(n - \bar{n})^a(n - \bar{n})^a}{u - \bar{u}} \right]\ , \\  
  G_{11} & = & \frac{2 \imath }{u - \bar{u}}\, \sqrt{G} \ ,
\qquad \qquad G_{12} = \imath\frac{ u + \bar{u}}{u - \bar{u}}\, \sqrt{G}\ .
\nn
\end{eqnarray}
In this new field basis the scalars of the vector multiplets 
are collectively denoted by $z^i = (s,t,u,n^a),\, a = 4,\ldots,n_V$.
Inserting (\ref{e11})
into (\ref{an2}) using (\ref{e1})
the action is found to be
\be
  \label{asgt}
  S =   2 \int d^4 x\sqrt{-g}  \Big[ \f R 
  + \frac{1}{4}\, I_{IJ} F^I_{\mu \nu} F^{J\mu \nu}  
  + \frac{1}{8} \, R_{IJ} F^I_{\mu \nu} F^J_{\rho\lambda}
  \epsilon^{\mu \nu \rho \lambda}   
   - g_{i\bar\jmath} {\partial}_\mu z^i {\partial}^\mu
  \bar{z}^{\bar\jmath} - h_{uv} {\partial}_\mu q^u {\partial}^\mu q^v 
  \Big] \ ,\nn 
\ee
where the metric $g_{i\bar{\jmath}}$ is K\"ahler, i.e.\
$g_{i\bar{\jmath}} = \partial_i\partial_{\bar\jmath} K$ with
\be\label{Khet}
 K = - \, \ln \imath (\bar{s} -s) 
-\,\ln{ \frac{1}{4}\left[(t - \bar{t})(u - \bar{u}) - (n - \bar{n})^a 
(n -\bar{n})^a  \right]} \ .
\ee
This is the K\"ahler potential of the coset (\ref{SKcoset}).

In the following we choose
to relabel the indices in order to be consistent with the standard
$N=2$ conventions. More specifically, the index $I$ which 
runs over all $n_V+1$ vector fields of the theory 
starts at $I=0$, i.e.\ $I=0,\ldots,n_V$ where $I=0$ denotes the graviphoton. 
Furthermore, the matrix $\eta$ defined in eq.\ (\ref{eta}) 
is relabeled such that $\eta_{01}=\eta_{10}=\eta_{23}=\eta_{32}= \eta_{aa}=1$
while all other matrix elements vanish. The matrix $M_{IJ}$ is 
relabeled accordingly and conventionally written as
\be
  \label{e13}
M_{IJ}\  =\
   \eta_{IJ} - 2\, \frac{(X_I \bar{X}_J + \bar{X}_I
  X_J)}{X^I \eta_{IJ} \bar X^J} \ ,
\ee
where the coordinates
$X^I, I=0,\ldots,n_V$
are related to the toroidal moduli via
\begin{equation}
  \label{e23}
  X^0 =  \frac{1}{2}  \, , \quad
  X^1 =  \frac{1}{2} (u t -n^an^a) \, , \quad
  X^2 = - \frac{1}{2} u \, , \quad
  X^3 =  \frac{1}{2} t\, , \quad
  X^a = \frac{1}{\sqrt 2}\, n^a\ ,
\end{equation}
and we defined  $X_I \equiv \eta_{IJ} X^J $.
In these coordinates the K\"ahler potential (\ref{Khet}) reads
\begin{equation}
  \label{Ks}
  K  = - \, \ln \imath (\bar{s} - s) - \ln{X^I\eta_{IJ} \bar X^J} \ .
\end{equation}

Note that the coordinates $X^I$ are not all independent but satisfy
the constraint
\begin{equation}\label{Xcon}
  X^I \eta_{IJ} X^J = 0 \ .
\end{equation}
This is related to the fact that in the heterotic string 
one naturally obtains a basis of field variables which
differ from the standard $N=2$ convention discussed in
appendix~C by a symplectic
rotation \cite{CAFP,dWKLL,AFGNT,LF,N2}. 
In the standard field variables the coset 
(\ref{SKcoset}) is described by a holomorphic prepotential 
\begin{equation}
F\  =\ {X^1(X^2X^3 - X^aX^a)\over X^0} \ .
\ee
By performing the symplectic rotation 
$X^1 \to -F_1,\ F_1 \to  X^1,$
one obtains the field basis used in (\ref{e23}) where
the constraint (\ref{Xcon}) holds and a holomorphic
prepotential does not exist \cite{CAFP,dWKLL,AFGNT,LF,N2}.

Now we are prepared to discuss the theory with
background fluxes turned on.
We perform the analysis in two steps.
First we compactify the ten-dimensional heterotic
theory on $T^2$ with all possible  fluxes turned on
and then do a further compactification on $K3$ where all
fluxes along the $K3$ vanish.
In a second step we switch off the fluxes along $T^2$ but now
consider all possible fluxes along the $K3$.
Finally we discuss the combined situation.
Certain aspects of this analysis have also been done in ref.\ \cite{CKKL}.

\subsection{Flux along $T^2$}
\label{ft2}

For the $T^2$ compactification with fluxes we can straightforwardly
use the the results of \cite{KM} which we reviewed
in section 2.2. For a two-torus the formuli considerably simplify.
First of all no $H$-fluxes can be turned on
since that needs at least three internal indices, i.e.\ a $T^3$. 
Furthermore in the gravitational sector an internal axion only starts
to appear in $d=8$, as we discussed in section \ref{fluxtn}. This means that
non-zero fluxes can only be turned on
for the field strength of the gauge fields
which are already present in $d=10$. Consequently, the only non-vanishing
structure constants which we obtain from eq.\ (\ref{fflux})  on $T^2$ are
$f^a_{12}= 2 m^a$. In the relabeled field basis introduced in the
previous section the non-vanishing
structure constants become\footnote{Notice that the indices are now raised and
lowered
with the `new' $\eta$-matrix and hence we have 
$ f^a_{30}=f^2_{0a}=f^1_{a3}$ .}
\begin{equation}\label{fluxT2}
f_{30}^a\  =\  2 m^a \ .
\end{equation}
Inserted into (\ref{potKM}) the potential dramatically simplifies and 
reads
\begin{equation}
  \label{e4.1}
  \V = 2 \det (G^{-1})\,  m^a m^a \ .
\end{equation} 
The compactification of the eight-dimensional version of the action (\ref{eq3})
on $K3$ proceeds exactly as in
the ungauged case and results in
\begin{eqnarray}
  \label{an2g}
  S & = & 2 \int d^4 x \sqrt{-g} \; \Big[ \f R 
+ \frac{1}{4}\, I_{IJ} F^I_{\mu \nu} F^{J\mu \nu}  
+ \frac{1}{8} \, R_{IJ} F^I_{\mu \nu} F^J_{\rho\lambda}
\epsilon^{\mu \nu \rho \lambda} \\  
&& \qquad - g_{i\bar\jmath} {\mathcal D}_\mu z^i {\mathcal D}^\mu \bar{z}^{\bar\jmath}
  - h_{uv} \partial_\mu q^u \partial^\mu q^v
- \VE  \Big] \ ,\nn 
\end{eqnarray}
where the covariant derivatives are given by\footnote{%
These covariant derivatives are nothing but the covariant
derivatives of eq.\ (\ref{potKM}) expressed in the coordinates 
(\ref{e11}) with the specific choice of structure constants given in
(\ref{fluxT2}).}
\begin{eqnarray}
  \label{e16}
  \mathcal{D}_\mu u & = & \partial_\mu u \, , \qquad
  \mathcal{D}_\mu s =  \partial_\mu s \, , \nn \\
 \mathcal{D}_\mu t & = &   \partial_\mu t - 2\sqrt 2 A_\mu^3 m^a n^a +
  2 m^a A^a_\mu \, , \nn \\ 
  \mathcal{D}_\mu n^a & = & \partial_\mu n^a -\sqrt 2 (A_\mu^3 u +
  A_\mu^0) m^a \, .
\end{eqnarray}
The potential $\VE$ is obtained from (\ref{e4.1}) by an appropriate
Weyl rescaling and 
inserting the coordinate transformations (\ref{e11}) one finds
\be
  \label{e28}
\VE\ \equiv\ \frac{-\imath \V}{2 (s - \bar{s})}\ =\
\,  \frac{ 4 \imath (u - \bar{u})^2\ m^a m^a}{(s - \bar{s})[(n-\bar{n})^2 -
  (u - \bar{u}) (t - \bar{t})]^2 }\ .  
\ee

We see that compared to the ungauged action (\ref{an2}) 
the fluxes induce a potential $\VE$ and turn some of the derivatives
of the scalars in the vector multiplets into covariant derivatives.
Thus a subgroup of the isometry group of the vector multiplets is gauged
by the presence of the fluxes $m^a$. 

We can check the consistency
of the low energy effective theory with the standard $N=2$ gauged 
supergravity which we briefly summarize in appendix~C.
The generic form of the covariant derivatives of the scalars 
is given by
\begin{equation}\label{zcov}
 \mathcal{D}_\mu z^i = \partial_\mu z^i - k_I^i A_\mu^I\ ,
\end{equation}
where the $k_I^i$ are the Killing vectors of the isometries 
$\delta z^i = \Lambda^I k_I^i$.
Comparing with  (\ref{e16})
we can read off the non-zero Killing vectors
\begin{equation}
  \label{e26}
  k_3^t = 2 \sqrt 2 m^a n^a \, , \qquad
     k_a^t =  -2 m^a \, ,\qquad
     k_0^{n^a}  =   \sqrt{2} m^a \, , \qquad  k_3^{n^a}  = 
    \sqrt{2} m^a u  \ . 
\end{equation}
In the case that only isometries in the vector multiplet sector
are gauged the general $N=2$ potential given in (\ref{pot})
simplifies and reads
\begin{equation}
  \label{e15}
  \VE =   e^{K} g_{i \bar{\jmath}} k^i_I k^{\bar{\jmath}}_J
  \bar{X}^I X^J \ .
\end{equation}
Inserting  (\ref{e26}) into (\ref{e15}) and using the definitions (\ref{e23})
one verifies the consistency with  the potential 
given in (\ref{e28}).

\subsection{Fluxes on $K3$}
\label{fK3}
Let us now discuss the case where non-trivial background fluxes 
along the $K3$ but not along $T^2$  
are turned on. On $K3$ there are 22 harmonic two-forms $\Omega_A$  and one
harmonic four-form $\hat\Omega_{(4)}$ but no three-form.
Hence, the possible fluxes can only arise from the gauge field strength
$F$
while the three-form $H$ cannot contribute. More precisely, we expand the 
(Abelian) gauge fields with vanishing instanton number or in other words
the gauge fields which are not involved in the solution of the constraint
(\ref{K3cons}) according to
\begin{equation}
  \label{e36}
  F^I = F^I(x) + m^{IA}\, \Omega_A\ ,\qquad A=1,\ldots,22\ ,
\end{equation}
where $F^I(x)$ is the field strength in the four-dimensional space-time
and $\Omega_A$ are the 22 harmonic two-forms on $K3$.
Furthermore, we insist that no hypermultiplet is charged
under the gauge fields for which $m^{IA}\neq 0$.\footnote{The case where
hypermultiplets are charged with respect to gauge fields with fluxes
is discussed in refs.\ \cite{Bac,AGNT}.}

Since we are turning on the fluxes perturbatively the constraint 
(\ref{K3cons}) is modified. However, we do not change the instanton 
configuration and continue to choose  part of the gauge group
with non-trivial instanton number,
i.e. $\int_{K3} tr F^2 = 24$. In order to satisfy (\ref{K3cons}) the 
additional fluxes $m^{AI}$ then have to obey
\begin{equation}
  \label{K3con}
  0 =  \int_{K3} \eta_{IJ} F^I \wedge F^J = m^{AI} m^{BJ}
  \rho_{AB} \, \eta_{IJ} \ ,
\end{equation}
where 
$\rho$ denotes the intersection matrix on the second cohomology $H^2(K3)$
\begin{equation}
  \label{defrho}
  \rho_{AB} = \int_{K3} \Omega_A \wedge \Omega_B\ .
\end{equation}
$\rho$ has signature $(3,19)$ and is left invariant by $SO(3,19)$.
Note that the condition (\ref{K3con}) is the equivalent of (\ref{T^ncon}) 
for toroidal compactifications. 

Due to the Chern-Simons couplings the action cannot be written only
in terms of $F$ and thus the consistency of the reduction is not entirely
obvious.
However, by dualizing the antisymmetric tensor $B$ in eight dimension to a 
four-form $C_4$ \cite{CKKL}
 the kinetic term $\int e^{-\phi} H_{\mu\nu\rho}^2$ 
turns into
\begin{equation}
  \label{e34}
  S_C = \int e^{\phi} G_5^2 - \int C_4 \wedge F^I \wedge F^J \eta_{IJ} \ ,
\end{equation}
where $G_5 = dC_4$. In this dual basis the Kaluza-Klein reduction
can be performed with $F^I$ given in (\ref{e36}) and the four-form 
$C_4$ expanded
according to
\begin{equation}
  \label{e35}
  C_4 = c^A(x) \wedge \Omega_A + a(x)\, \hat\Omega_{(4)} \ .
\end{equation}
The $c^A$ are 22 two-forms in space-time while
$\hat\Omega_{(4)}$ is the volume form on $K3$ and $a(x)$ the associated 
scalar field.\footnote{The space-time part of $C_4$ does not contribute
since it does not have a kinetic term and a possible contribution to
the potential vanishes due to (\ref{K3con}).}
After the reduction one can dualize the 22
antisymmetric tensors $c^A$  into axionic scalar fields $b^A$.
($a(x)$ is the axion which arises from dualizing
$B_{\mu\nu}$ in $d=4$.) The final action in the $d=4$ 
Einstein frame reads
\begin{eqnarray}
  \label{an2.1}
  S & = & 2 \int d^4x \sqrt{-g}\Big[ \f R 
  + \frac{1}{8}\, I_{IJ} F^I_{\mu \nu} F^{J\mu \nu}  
  + \frac{1}{16} \, R_{IJ}F^I_{\mu \nu} F^J_{\rho\lambda}\epsilon^{\mu \nu
    \rho \lambda} \\ 
  & & - g_{i\bar j} \partial_\mu z^i \partial^\mu \bar z^{\bar j}
  - h_{uv} {\mathcal D}_\mu q^u {\mathcal D}^\mu q^v
  - \VE
  \Big] \ ,\nn 
\end{eqnarray}
where among the hypermultiplet scalars $q^u$ only the
22 axions $b^A$ are charged and their covariant derivatives are given by
\begin{equation}
  \label{e38}
  {\mathcal D}_\mu b^A =  \partial_\mu b^A - m_I^A A_\mu^I \ .
\end{equation}
The potential is found to be\footnote{For a subset of fluxes
and at special points in the $K3$ moduli space 
this potential has also been obtained in ref.\ \cite{CKKL}.}
\begin{equation}
  \label{e41}
  \VE = \frac{2}{(s - \bar{s})^2}\  h_{AB}\, m^{AI} m^{BJ}\, I_{IJ}\ ,
\end{equation}
where $h_{AB}$ is the restriction of the quaternionic metric $h_{uv}$ to the
space spanned by $b^A$ and is given by:
\begin{equation}
  \label{e37}
  h_{AB} =  \frac{1}{4 v} \, N_{AB} \, , \quad N_{AB}(q) = \int_{K3}
  \Omega_A \wedge ^* \! \Omega_B \ . 
\end{equation}
Here $v$ is the volume modulus of $K3$ while $N_{AB}$ depends only on the
remaining 57 moduli of $K3$ and not on the 22 axions $b^A$.
From (\ref{e38}) we see that contrary to the previous case (fluxes along $T^2$)
the non-trivial fluxes along $K3$ gauge isometries in the hypermultiplet 
moduli space ${\cal M}_H$.

As before the final step is to check the consistency with
the standard gauged supergravity. Inserting (\ref{IM}) into (\ref{pot}) 
using (\ref{e13}) and $k_I^i=0$ one
finds that the gauging in the hypermultiplet sector induces the potential
\begin{equation}
  \label{potK3}
  \VE \ =  \ 4 e^K X^I \bar X^J h_{uv} \, k^u_I k^v_J - \frac{\imath}{s 
    - \bar s} \; \eta^{IJ} \, P_I^x P_J^x \ ,
\end{equation}
where the $P_I^x, x= 1,2,3$ are the Killing prepotentials defined in appendix~C.

For the present case where only the axionic shift (Peccei-Quinn) 
isometries $b^A \to b^A +
m^A_I \Lambda^I$ are gauged, the Killing prepotentials can be computed
explicitly on the subspace ${\cal M}_{K3}$ given in (\ref{K3sub})
which is  spanned by the K3 moduli. 
The detailed calculation is presented in appendix \ref{apD} 
with the final result 
\begin{equation}
  \label{px^2}
  \eta^{IJ} P^x_I P^x_J = \eta^{IJ} \big( h_{uv} k^u_I k^v_J - m^A_I m^B_J
  \rho_{AB} \big) = \eta^{IJ} h_{uv} k^u_I k^v_J \, ,
\end{equation}
where the last equation holds due to the constraint (\ref{K3con}).
Inserted into (\ref{potK3}) one obtains
\begin{equation}
  \label{e40}
  \VE \ = \  \frac{\imath}{s - \bar s} \ h_{uv} k^u_I k^v_J \Big[ 2 \ 
  \frac{X^I \bar X^J + \bar X^I  X^J}{X^K \bar X_K} - \eta^{IJ} \Big]
 \ = \  \frac{2}{(s - \bar s)^2} \,  h_{uv} k^u_I k^v_J I^{IJ} \ ,
\end{equation}
where we used (\ref{IM}),(\ref{e13}) and (\ref{Ks}).
With the killing
vectors which can be read off from (\ref{e38}) 
$ k^u_I = m_I^A$, one immediately shows the equivalence of the potential
derived from compactification (\ref{e41}) and the potential
obtained from gauged supergravity (\ref{e40}).

Note that in the computation just described we had to set the 
moduli fields arising from the gauge bundle to zero.
Only in this case we are able to compute the $P_I^x$.
However, the consistency of the gauged supergravity with
the compactification suggests that including the bundle
moduli does not alter the $P_I^x$. It would be interesting
to verify this explicitly.

\subsection{Fluxes simultaneously on $T^2$ and K3}

Let us now analyze the case where fluxes along both $T^2$ and $K3$ 
are non-vanishing. We have seen in the previous two sections that the only
candidates for turning on a background value are the field strengths of the
vector fields. The consistency of the procedure was guaranteed by the Abelian
gauge invariance which in turn was obtained by going to the Coulomb branch of
the theory. On the other hand we have shown in section \ref{fluxtn} that 
turning on fluxes in the 
torus compactifications generates the gauging of a non-Abelian group. Naively
this would imply that no fluxes can be turned on in a further compactification
on $K3$. 
However, for the vector fields arising from the
reduction of the antisymmetric tensor $B$ in ten dimensions
non-trivial fluxes can be turned on as these vector fields appear
in the action only  via their Abelian field strength. This is a 
direct consequence of the two-form gauge invariance 
$B\to B + d\Lambda$ in $d=10$.
 Moreover if some of the fluxes in
(\ref{fluxT2}) vanish, 
the corresponding gauge fields also only appear via their Abelian
field strength and hence 
it is possible to turn on fluxes along $K3$ 
as in eq.\ (\ref{e36})
for all the gauge fields in $d=8$ which have an Abelian symmetry.
The derivation of the effective action proceeds exactly as before
with the result
\begin{eqnarray}
  \label{Ssim}
  S & = & 2 \int \sqrt{-g} \; d^4 x \Big[ \f R 
+ \frac{1}{8}\, I_{IJ} F^I_{\mu \nu} F^{J\mu \nu}  
+ \frac{1}{16} \, R_{IJ}F^I_{\mu \nu} F^J_{\rho\lambda}\epsilon^{\mu \nu
\rho \lambda} \\ 
&& - g_{i\bar j} {\mathcal D}_\mu z^i  {\mathcal D}^\mu \bar z^{\bar j}
  - h_{uv} {\mathcal D}_\mu q^u {\mathcal D}^\mu q^v
- \frac{\imath\V}{2 (s - \bar{s})}  \Big] \ ,\nn 
\end{eqnarray}
where the structure of the covariant derivatives is exactly as in 
(\ref{e16}) and (\ref{e38})
and the potential is the sum of the potentials
given in eqs.\ (\ref{e28}), (\ref{e41}).
Since also in the gauged supergravity the potential (\ref{pot})
is a sum of two terms the consistency with the reduction
follows immediately from our previous analysis.

\subsection{Compactification on $K3\times S^1$}
Finally, let us briefly discuss the compactification of
the heterotic string on $K3\times S^1$ to $d=5$. 
The derivation of the effective theory
is very similar to the previous case and therefore we only highlight
the differences.
The five-dimensional gauged supergravity (with 8 supercharges)
again contains a gravitational multiplet, $n_V$ vector multiplets
and $n_H$ hypermultiplets. Now the scalars
in the vector multiplets are real and their scalar manifold
is a so called `very special' manifold \cite{GST} while 
the hypermultiplets including their quaternionic geometry remain
unchanged.

The effective theory of the the heterotic string compactified on $K3\times
S^1$
without fluxes has been derived in ref.\ \cite{AFT}. 
Including background fluxes again proceeds in two steps.
First one compactifies the ten-dimensional heterotic string
on an $S^1$ to $d=9$. In this step no fluxes
can be turned on. Only by further compactification on $K3$ 
all fluxes as in (\ref{e36}) can be chosen non-vanishing.
Thus in the resulting five-dimensional theory
the vector multiplet sector remains ungauged and only the scalar fields in
the hypermultiplet become charged.  
This is in full agreement with the
results of gauged five-dimensional supergravity as obtained in 
refs.\ \cite{GST,LOSW,GZ,CD}. 
Furthermore, the covariant derivatives are exactly as in (\ref{e38})
and the scalar potential is found to be
\begin{equation}
\label{potd5}
\VE = \frac1s\  h_{AB}\, m^{AI} m^{BJ}\, M_{IJ} \ ,
\end{equation}
where $M$ is the matrix (\ref{e1}) for the case where only one dimension is
compactified on a circle. Its relation to the gauge couplings is
\begin{equation}
  M_{IJ}\ =\ \frac2s\, I_{IJ} \ , \qquad s \ \equiv\ e^{-\frac23 \phi}\ .
\end{equation}
Note that we are using the basis where the antisymmetric tensor field $B_{\mu
  \nu}$ and not the dual gauge field  appears explicitly in the action. 
The low energy effective action is described in terms of the
derivatives of a cubic 
prepotential $V$
subject to the constraint $V=1$. For the heterotic string compactified on
$K3\times S^1$ the scalars in the vector multiplets
span the coset $SO(1,n_V)/SO(n_V)$ and the corresponding
prepotential reads \cite{AFT}
\be\label{F5d}
V = S X^I X^J \eta_{IJ} \ ,
\ee
where the $X^I$ are now real.
The gauge couplings  are given by 
\begin{equation}
  \label{gc5d}
  I_{IJ} = \partial_I \partial_J \ln V \vert_{V=1} \ .
\end{equation}

The consistency with the gauged supergravity of \cite{GST,LOSW,GZ,CD}
can be established using  exactly same steps as in section
\ref{fK3} and so we will not present this in detail. We only note that the
scalar potential of the $N=2$, $d=5$ gauged supergravity has a form similar to
the one given in the four-dimensional case 
\begin{equation}
  \label{pk3s1}
  \VE = (2 I^{IJ} - 4 X^I X^J) P^x_I P^x_J + 4 X^I X^J h_{uv} k^u_I k^v_J \, , 
\end{equation}
where $I^{IJ}$ are the inverse gauge couplings. 
From (\ref{F5d}) and (\ref{gc5d}) we learn 
\begin{equation}
  \label{IIJ}
I^{IJ} =  \frac{1}{s^2} \, \eta^{IK} \eta^{JL} I_{KL}   = - \frac1s \
  \eta^{IJ} + 2 X^I X^J \ .
\end{equation}
The geometry of the  hypermultiplets is identical in $d=4,5,6$ 
and as a consequence the Killing vectors and also the computation of 
the Killing prepotentials $P^x_I$ presented in appendix D is unaltered.
Using (\ref{pxpx}), (\ref{K3con}) and (\ref{IIJ}) one then verifies
the consistency of the potentials (\ref{pk3s1}) and (\ref{potd5}).

\setcounter{equation}{0}\section{Conclusions}

In this paper we studied the compactification of the heterotic string to 
$d=4, \,  5$ with 8 or 16 preserved supercharges in the presence of
background fluxes. 
We showed that in each case the bosonic sector of the low energy effective
action is consistent with the
corresponding gauged supergravity. 
However, the induced
potential displays a dilaton instability and thus 
no Minkowski or anti-de Sitter ground state exists.

Interestingly, by turning on fluxes we were not able to recover
all possible couplings of 
the gauged supergravities.
In the case of 16 supercharges the gauged supergravity can have 
a constant term in the (string-frame) potential while in
toroidal compactifications of the heterotic string this term is absent.
Furthermore, in $d=5$ it is possible to gauge
an Abelian group where the dual of the heterotic antisymmetric tensor field 
plays the role of the gauge field \cite{DHZ}.
Also this type of gauging could not be obtained by turning on
background fluxes. 

In supergravities with 8 supercharges one can gauge the isometries of 
the vector- and hypermultiplet geometries. On the other hand,
in $K3$ compactifications of the heterotic string only 
a small subset of the isometries is gauged and in most cases these
are axionic shift (Peccei-Quinn) symmetries.

For toroidal compactifications of the heterotic string in $d>5$
the bosonic action is also known but unfortunately the
corresponding gauged supergravities have not been constructed
yet and a hence a comparison cannot be performed.
Finally, in compactifications of the heterotic string
on $K3$ to $d=6$ no background fluxes can be turned on at all
since there are no Abelian gauge fields in $d=10$.

\newpage
\appendix
\noindent
{\Large {\bf Appendix}}
\renewcommand{\theequation}{\Alph{section}.\arabic{equation}}
\setcounter{equation}{0}
\setcounter{section}{0}
\section{Conventions and notations}
Let us assemble our conventions in this appendix.
The space time metric is taken to have the signature: $(-,+,+, \ldots +)$
while the Riemann tensor is defined to be $R^\mu_{\nu\lambda\rho} =
\partial_\lambda \Gamma^\mu_{\nu\rho} 
- \partial_\nu \Gamma^\mu_{\lambda\rho}
+ \Gamma^\mu_{\lambda\sigma} \Gamma^\sigma_{\nu\rho} - \Gamma^\mu_{\nu\sigma} \Gamma^\sigma_{\lambda\rho}$.

Throughout the paper we try to use the same indices
for similar quantities but their range does change in the various
compactifications. More specifically we use 
(if not specified otherwise)

\begin{equation}
  \label{notation}
  \begin{array}{rcll}
    \mu,\ \nu \ & = &  0, \ldots, d-1 \hspace{2cm} & \textrm{label Minkowskian
    space-time indices} \\ 
    \alpha, \ \beta  & = & 1, \ldots, n&\textrm{label directions of}\ T^n \\ 
    a, \ b & = & 1, \ldots, 16 & \textrm{label 
      the Cartan subalgebra of the  }\\ &&&\textrm{heterotic gauge group}\\
    I, \ J & = & 1, \ldots, 16 + 2n & \textrm{label all gauge fields} \\
    i,\ j & = & 1, \ldots, n_V & \textrm{label all vector multiplets} \\
    A,\ B & = & 1, \ldots, 22 & \textrm{label the harmonic 2-forms on K3} \\
    \Lambda, \ \Sigma & = & 1, \ldots, 4 & \textrm{label } USp(4)\ 
    \textrm{in } T^5\textrm{ reduction and } \\ & & & SU(4)\
     \textrm{in } T^6 \textrm{ reduction}. \\ 
  \end{array}
\end{equation}
$g_{\mu \nu}$ denotes the space-time metric, while $G_{\ax \bx}$ is used for
the metric on the internal space.

\section{$N=4$ gauged supergravity in $d=5$}
\label{N4d5}
\setcounter{equation}{0}

The purpose of this appendix is to supply some of the details
of the computation outlined in section 2.3.  
Let us first recall some basic facts about  $USp(4)$.
It has an invariant symplectic two-form $\Omega$ which satisfies
\begin{equation}
  \Omega_{\Sx \Dx} \Omega^{\Dx \Lx} = - {\Omega_\Sx}^\Dx {\Omega_\Dx}^\Lx = -
  {\delta_\Sx}^\Lx \ ,
\end{equation}
where
the indices $\Sx,\Lx = 1,\ldots, 4$ are raised and lowered with $\Omega$
and the convention for the matrix product is
\begin{equation}
  \label{mult}
  {\left(A B\right)_\Sx}^\Lx = {A_\Sx}^\Dx {B_\Dx}^\Lx \ .
\end{equation}

The other quantities we need in the main text are
the five Euclidean 
$4\times4$ $\Gamma$-matrices satisfying 
\begin{equation}\label{GammaM}
  {\left\{ \Gamma^\ax, \Gamma^\bx \right\}_\Sx}^\Lx = 2\, \delta^{\ax \bx}
  {\delta_\Sx}^\Lx \, , \qquad  Tr \left( \Gamma^\ax \Gamma^\bx \right) 
= 4\, \delta^{\ax  \bx}\ , \quad \alpha=1,\ldots,5\ .
\end{equation}
Furthermore, the $\Gamma$-matrices are chosen
to obey\footnote{See appendix B of ref.\ \cite{DHZ}.}
\begin{equation}
  \label{cstg}
  \left( \Gamma^\ax \right)_{\Sx \Lx} = - \left( \Gamma^\ax \right)_{\Lx \Sx} 
\quad
\textrm{and} \quad \left( \Gamma^\ax \right)_{\Sx \Lx} \Omega^{\Sx \Lx} = 0\ .
\end{equation}

In order to relate the coordinates $L^I_{\Sx \Lx}$ to the matrix $M^{IJ}$ used
in \cite{KM} one defines the quantities $Z_\ax^I$ by
\begin{equation}
  \label{dfz}
Z_\alpha^I = \frac{1}{2} {\left( \Gamma^\ax\right)_\Lx}^\Sx {{L^I}_\Sx}^\Lx \
  , \qquad
  {{L^I}_\Sx}^\Lx = \frac{1}{2} {\left(\Gamma^\ax\right)_\Sx}^\Lx Z^I_\ax \ ,
\qquad I=1,\ldots,26\ .
\end{equation}
Inserted into (\ref{sccon}) and using (\ref{cstg})
one derives the constraint
on the $Z_\ax^I$ to be
\begin{equation}
  \label{cstzapp}
  Z_\ax^I Z_\bx^J \eta_{IJ} = - \delta_{\ax \bx}\ .
\end{equation}

After eliminating ${L_{i}}^I$ via the constraint
given in (\ref{sccon}) one typically
encounters terms of the form $Tr(L^{I_1} \ldots L^{I_k})$ where the trace and
the matrix multiplication is to be understood as acting on the indices $\Sx,
\Lx$ according to (\ref{mult}).
This can be easily computed 
using the trace properties of the $\Gamma$ matrices. As an example
let us consider $k=2$. Using (\ref{GammaM}) one derives
\be
  \label{zij}
  {L_I}^{\Sx \Lx} L_{J\Sx \Lx} =  \frac{1}{4} Z_{\ax I} Z_{\bx J}
  (\Gamma^\ax)^{\Sx \Lx} (\Gamma^\bx)_{\Sx \Lx} 
=   Z_{\ax I} Z_{\bx J} \delta^{\ax \bx}  \ \equiv \ Z_{IJ}\ .
\ee

Finally, we dualize the spectator gauge field $a_\mu$
to an antisymmetric tensor $B_{\mu\nu}$.
The terms in the Lagrangian (\ref{eq:act5.1}) which contain the field $a$ are
(written in form notation)
\begin{equation}\label{Sdual}
  \frac{1}{2}\int\left[- e^\phi G \wedge ^* \!G +  \eta_{IJ}
  F^I \wedge F^J \wedge a \right] \,
\end{equation}
where $F^I = d A^I - \frac{1}{2} f_{JK}^I A^J \wedge A^K$. 
The last term in (\ref{Sdual}) can be rewritten in terms
of the Chern-Simons term given in (\ref{FHdefmod})
\begin{eqnarray}\label{CS}
\eta_{IJ}  F^I \wedge F^J  &=& d \omega^{\rm CS}\\
\omega^{\rm CS}  &  =  &
\eta_{IJ} A^I \wedge F^J + \frac{1}{6} f_{IJK} A^I \wedge A^J
  \wedge A^K \ .\nonumber
\end{eqnarray}
Using (\ref{CS}) the equation of motion for $a$ becomes
\begin{equation}
  d(e^{\phi}{\,} {^* G} - \frac{1}{2} \omega_{cs}) = 0 \ .
\end{equation}
We introduce a two-form $B$ and its three-form field strength $H$ 
such that
\begin{equation}
  dB = -e^{\phi} {\,} {^*G} + \frac{1}{2} \omega_{cs} \ ,\qquad
  H = dB - \frac{1}{2} \omega_{cs}\ .
\end{equation}
The action for $H$ is now given by 
$ \frac{1}{2}\int e^{-\phi} H \wedge ^* \! H$ 
and we see that this is exactly the right kinetic term for $B$ as it 
is obtained from the compactification  in eq.\ (\ref{eq3}).

\section{$N=2$ gauged supergravity in $d=4$}
\setcounter{equation}{0}
\label{sg}
The purpose of this appendix is to give a short review of the $N=2$ 
supergravity in four dimensions \cite{wp,bw,N2}. 
A generic spectrum contains the gravitational multiplet,
$n_V$ vector multiplets and $n_H$ hypermultiplets.
The vector multiplets contain $n_V$ complex scalars
$z^i, i=1,\ldots,n_V$ while the hypermultiplets
contain $4n_H$ real scalars $q^u, u=1,\ldots, 4n_H$.
Due to supersymmetry the scalar manifold factorizes
\begin{equation}\label{modulispace}
\cM=\cM_V \otimes \cM_H \ ,
\end{equation}
where the component
$\cM_V$ is a  special K\"ahler manifold
spanned by the scalars $z^i$ while
$\cM_H$ is a quaternionic manifold spanned by the scalars $q^u$.

A special K\"ahler manifold is  a K\"ahler manifold  whose geometry 
obeys  an additional constraint \cite{wp}.
%
This constraint states that the 
 K\"ahler potential $K$ is not an arbitrary real function 
but 
determined in terms of a holomorphic prepotential $F$
according to
%
\begin{equation}\label{Kspecial}
K=-\ln\Big(i \bar{X}^{I} (\bar z^i) F_{I}(X)
- i X^{I}  (z)\bar{F}_{I}(\bar{X})\Big) \ .
\end{equation}
The $X^{I}, I=0,\ldots, n_V$ are $(n_V+1)$
holomorphic functions of the 
$z^i$.
$F_{I}$ abbreviates the derivative, i.e.
$F_{I}\equiv \frac{\partial F(X)}{\partial X^{I}} $
and 
$F(X)$ is  a homogeneous function of $X^{I}$ of degree $2$, i.e.\
$X^{I} F_{I}=2 F$.

The $4n_H$ scalars $q^u, u=1,\ldots,4n_H$ in the hypermultiplets are
coordinates on a quaternionic manifold \cite{bw}.
This implies the existence of three almost complex
structures $(J^x)^w_v, x=1,2,3$ which satisfy the
quaternionic algebra
\begin{equation}\label{Jx}
J^x J^y = -\delta^{xy} + i \epsilon^{xyz} J^z\ .
\end{equation}
Associated with the complex structures is a
triplet of K\"ahler forms
\begin{equation}\label{Kx}
K^x_{uv} = h_{uw} (J^x)^w_v\ ,
\end{equation}
where $h_{uw}$ is the quaternionic metric.
The holonomy group of a quaternionic manifold
is $Sp(2)\times Sp(2n_h)$ and
$K^x$ is identified with the field strength of the
$Sp(2)\sim SU(2)$ connection $\omega^x_v$, i.e.
\begin{equation}\label{Kdef}
K^x \ = \
d\omega^x +\frac12\,\epsilon^{xyz}\omega^y\wedge \omega^z~.
\end{equation}

The bosonic part of the $N=2$ action is given
by 
\be
  \label{asg}
  S =   \int d^4 x\sqrt{-g}  \Big[ \f R 
  + \frac{1}{4}\, I_{IJ} F^I_{\mu \nu} F^{J\mu \nu}  
  + \frac{1}{8} \, R_{IJ} F^I_{\mu \nu} F^J_{\rho\lambda}
  \epsilon^{\mu \nu \rho \lambda}   
   - g_{i\bar\jmath} {\partial}_\mu z^i {\partial}^\mu
  \bar{z}^{\bar\jmath} - h_{uv} {\partial}_\mu q^u {\partial}^\mu q^v 
  \Big] \ ,\nn 
\ee
where $g_{i\bar\jmath}=\partial_i\partial_{\bar\jmath} K$
and the gauge coupling functions are given by 
\bea
I_{IJ} &\equiv& {\rm Im} {\cal N}_{IJ}\ ,\qquad
 R_{IJ}\equiv {\rm Re} {\cal  N}_{IJ}\ , \nn\\
{\cal N}_{IJ} &=& \bar F_{IJ} +2i\ {\mbox{Im} F_{IK}\mbox{Im} F_{JL} X^K X^L
\over \mbox{Im} F_{LK}  X^K X^L} \ .
\eea
$F^0_{\mu\nu}$ denotes the field strength of
the graviphoton.

One can gauge 
the isometries on the scalar manifold ${\cal M}$.
Such isometries are generated by the
Killing vectors $k_I^u(q),\, k_I^i(z)$
\begin{equation}\label{kdef}
\delta q^u \ = \ \Lambda^I k_I^u(q) \  ,\qquad
\delta z^i \ = \ \Lambda^I k_I^i(z) \  .
\end{equation}
$k_I^u(q),\, k_I^i(z)$ satisfy the Killing equations
which in $N=2$ supergravity can be solved in terms of
four Killing prepotentials $(P_I,P_I^x)$.
The Killing vectors on $\cM_V$ are holomorphic
and obey
\begin{equation}
k_I^i(z) = g^{i\bar j} \partial_{\bar j} P_I\ ,
\end{equation}
while 
the Killing vectors on $\cM_H$ are determined by a
triplet of Killing prepotentials $P_I^{x}(q)\,$
via
\begin{equation}\label{killingpre}
k^u_I\,K_{uv}^x \ = \  -D_v P_{I}^x \equiv
- (\partial_v P_{I}^x
+ \epsilon^{xyz} \omega_{v}^y P_I^z)\ .
\end{equation}
Gauging the isometries (\ref{kdef}) requires the replacement of 
ordinary derivatives by covariant derivatives in the action
\begin{equation}\label{gaugeco}
\partial_\mu q^u \to {\mathcal D}_\mu q^u = \partial_\mu q^u - k_I^u A_\mu^I\ ,
\qquad
\partial_\mu  z^i \to 
{\mathcal D}_\mu  z^i = \partial_\mu  z^i - k_I^i A_\mu^I\ .
\end{equation}
Furthermore the potential
\begin{eqnarray}
  \label{pot}
  \VE  & = & \ e^K \Big[ X^I\bar X^J (g_{\bar \imath j}\, k_I^{\bar\imath} k_J^j +
  4h_{uv}\,k^u_Ik^v_J) + g^{i\bar \jmath} D_i X^I D_{\bar j} 
  \bar X^J P_I^x P_J^x - 3 X^I \bar X^J P_I^x P_J^x \Big] \nn\\
  & = & e^K X^I \bar X^J (g_{\bar \imath j} \, k_I^{\bar\imath} k_J^j + 4h_{uv}\, k^u_I k^v_J) -
  \Big( \frac12 I^{-1 IJ} + 4 e^K X^I \bar X^J \Big)
 P_I^x P_J^x \ ,  
\end{eqnarray}
has to be added to the action in order to preserve supersymmetry.
The bosonic part of the action of gauged $N=2$ supergravity is 
then given by 
\begin{eqnarray}
  \label{agsg}
  S & = & 2 \int  d^4 x\sqrt{-g} \; \Big[ \f R 
  + \frac{1}{4}\, I_{IJ} F^I_{\mu \nu} F^{J\mu \nu}  
  + \frac{1}{8} \, R_{IJ} F^I_{\mu \nu} F^J_{\rho\lambda}
  \epsilon^{\mu \nu \rho \lambda} \nn\\  
  && - g_{i\bar\jmath} {\mathcal D}_\mu z^i {\mathcal D}^\mu
  \bar{z}^{\bar\jmath} - h_{uv} {\mathcal D}_\mu q^u {\mathcal D}^\mu q^v 
  - \VE  \Big] \ .
\end{eqnarray}

\section{Computation of the Killing prepotential $P^x_I$}
\label{apD}
\setcounter{equation}{0}

In this appendix we compute explicitly the Killing 
prepotentials $P_I^x$ for the specific gauging which appears in section
\ref{fK3}. To do this, one needs 
to solve the differential equation (\ref{killingpre}).
Since we do not know the quaternionic manifold ${\cal M}_H$ for
generic compactifications the $P_I^x$ cannot be easily determined.
However, if we set to zero the moduli fields which arise from
the gauge bundle, the hypermultiplet moduli
space reduces to the coset ${\cal M}_{K3} = SO(4,20) / SO(4) \times SO(20)$.
In this case one can use the general formulas for homogenous spaces
given in \cite{N2,FGPT}.
 More specifically for the $SU(2)$  connection $\omega^x$ one has 
\begin{equation}
  \label{con}
  \omega^x = - \frac12\, Tr (\theta \ \Sx^x)\ ,
\end{equation}
where
$\theta$ is a $4\times 4$ matrix of one-forms which
in terms of the coset representatives $Z$ reads
\begin{equation}
  \label{th}
  \theta_{\ax \bx} = Z_{\ax I} \eta^{IJ} d Z_{\bx J} \; ,
\quad \ax,\bx =1,\ldots,4\ .
\end{equation}
The $\Sx^x$ 
are the three self dual 't Hooft matrices as given in
\cite{N2}. Note that the quantities $Z$ are precisely the ones introduced in
(\ref{dfz}) for the coset $SO(5,m) / SO(5) \times SO(m)$. However, from the
compactification on $K3$ one does not obtain directly the coset
representatives $Z$, but a $SO(4,20)$ matrix $\M^{-1}$ which should satisfy a
condition similar to (\ref{MZ}).
From the discussions in section \ref{fK3} it is clear that the
scalars coming from the reduction of the antisymmetric two-form $B$ play a
crucial role in the gauged theory as they are the only scalars which become
charged when fluxes are turned on along $K3$. That is why we are going to
choose the following parameterization of 
the matrix $\M^{-1}$
 
\begin{equation}
  \label{M-1}
  \M^{-1} = \left( 
    \begin{array}{ccc}
      v^{-1} & - v^{-1} b^T \eta & - \frac12 v^{-1} (b^T \eta \, b) \\
      & & \\
      - v^{-1} \eta \, b &  v^{-1} \eta \, b \, b^T \eta + N & \frac12
      v^{-1} (b^T \eta \, b) \, \eta \, b +  N b \\
      & & \\
      - \frac12 \, v^{-1} (b^T \eta \, b) & \frac12 \, v^{-1} (b^T \eta \, b)
      \, b^T \eta + b^T N & \frac14 \, v^{-1} (b^T \eta \, b)^2 +  b^T N b + v
      \\ 
    \end{array} \right) \, ,
\end{equation}

\noindent
which precisely singles out the $b$'s while the dependence on the other
57 moduli of $K3$ is only via the $SO(3,19)$ matrix $N$
defined in (\ref{e37}). ($v$ is the volume of the $K3$.)
The matrices $\M$ and
$N$ are elements of $SO(4,20)$ and $SO(3,19)$  respectively
in that they obey 
\begin{equation}
  \M^{-1} \L \M^{-1T} = \L \, , \qquad N \, \rho \, N^T = \rho \ .
\end{equation}
$\L$ and $\rho$ are $SO(4,20)$ respectively $SO(3,19)$ metrics given by
\begin{equation}
  \L = \left(
  \begin{array}{ccc}
    0 & 0 & 1 \\
    0 & \rho & 0 \\
    1 & 0 & 0 \\
  \end{array} \right) \, , \qquad \rho_{AB} = \int_{K3} \Omega_A \wedge
    \Omega_B  \ .
\end{equation}

Let us define the $4 \times 24$ matrix
\begin{equation}
  \label{dZ}
  Z = \left(
    \begin{array}{ccc}
      \frac{1}{\sqrt 2} \, v^{-\frac12} & -\frac{1}{\sqrt 2} \,
      v^{-\frac12} \, b^T \eta & - \frac{1}{2 \sqrt 2} \, v^{-\frac12} 
      (b^T \eta \, b)  - \frac{1}{\sqrt 2} \, v^{\frac12}\\ 
      0 & u & u \, b \\
    \end{array} \right) ,
\end{equation}
where $u$ is a $3 \times 22$ matrix satisfying the analog of (\ref{MZ})
\be\label{uprop}
2 u^T u = N - \rho\ .
\ee
Furthermore $u$ depends only on the 57 $K3$ moduli but not on $b$.
It is easy to check that the matrix $Z$ defined in this way satisfies
\begin{equation}
  2 Z_{\ax I} Z_{\bx J} \delta^{\ax \bx} = \M^{-1}_{IJ} - \L_{IJ}\  .
\end{equation}
Inserting the coset representative $Z$ given in (\ref{dZ}) into (\ref{th}) 
the $4 \times 4$ matrix
$\theta$ is found to be
\begin{equation}\label{tf}
  \theta = \left(
    \begin{array}{cc}
      0 & \frac{1}{\sqrt 2} \, v^{-\frac12} d b^T u \\
      - \frac{1}{\sqrt 2} \, v^{-\frac12} u \, d b & u \, \rho \, d u^T \\
    \end{array} \right) \, .
\end{equation}
In order to determine the $P^x_I$ via eq.\ (\ref{killingpre})
we only need the $SU(2)$ connection $\omega^x$ in the direction
of the 22 axionic fields $b^A$. 
Inserting (\ref{tf}) into (\ref{con}) one finds
\begin{equation}
  \label{omegab}
  \omega_A^1\ =\ \frac{u_{A}^1}{\sqrt{2v}} \, , \qquad \omega_A^2 = -
  \frac{u_{A}^2}{\sqrt{2v}}   \, , \qquad 
\omega_A^3 = \frac{u_A^3}{\sqrt{2v}}   \, .  
\end{equation}

The solution of (\ref{killingpre}) is possible even if an explicit form of 
the matrix $u$ is not given. This is due to the fact that 
$\omega_a^x$  does not depend on $b$ and that the Killing
vectors point only in the $b$ direction.
Using (\ref{Kdef}), $k_I^u = m_I^A$ and $\partial_A \omega_v^x = 0$
we find
\be
  k_I^u K^x_{uv}  =  m_I^A \big(\partial_A \omega_v^x - 
  \partial_v \omega_A^x + \epsilon^{xyz} \omega_A^y \omega_v^z \big) 
  =  - D_v (m_I^A \omega_A^x) =  - D_v (k_I^u \omega_u^x) ,
\ee
which in turn implies that (\ref{killingpre}) is solved by 
\begin{equation}
  \label{px}
  {P}_I^x = k_I^u \omega_u^x\ .
\end{equation}

The term $P^x_I
P^x_J$ which appears in the $N=2$ potential (\ref{pot}) now yields
\be
  \label{pxpx}
  P^x_I P^x_J =  k_I^u k_J^v \omega^x_u \omega^x_v 
  =  \frac12 v^{-1}  u^x_{A}u^x_{B} m^A_I m^B_J =  \frac14
  v^{-1} \left(N_{AB} - \rho_{AB} \right) m^A_I m^B_J \ ,
\ee
where we used  (\ref{uprop}) and (\ref{omegab}). 
Inserting (\ref{e37}) we finally obtain
\be
P^x_I P^x_J  =  \left(h_{AB} - \frac{\rho_{AB}}{4 \, v} \right) m^A_I m^B_J =
  h_{uv} k^u_I k^v_J - \frac{\rho_{AB} m^A_I m^B_J }{4 \, v} \, .
\ee
Using (\ref{K3con}) this further simplifies and we have
\begin{equation}
  \eta^{IJ} P_I^x P_J^x = h_{uv} k_I^u k_J^v \eta^{IJ}\ .
\end{equation}

\vskip 1cm

{\large \bf Acknowledgments}

This work is supported by
the DFG (German Science Foundation),
GIF (German--Israeli
Foundation for Scientific Research),
the European RTN Program HPRN-CT-2000-00148
 and the
DAAD (German Academic Exchange Service).

We thank  C.\ Bachas, E.\ Bergshoeff, B.~Gunara, M.~Haack, C.~Herrmann, A.~Klemm, 
D.~L\"ust, T.~Mohaupt, H.~Singh, S.~Theisen and  M.~Zagermann
for useful conversations.



\end{document}